\documentstyle[psfig,a4wide,aps]{revtex} 
\newtheorem{gluck} { Theorem}
\newtheorem{laman} [gluck]{ Theorem}
\newtheorem{generic} [gluck] {Theorem}
\newtheorem{lemma} [gluck] {Lemma}
\begin{document}
\title{An efficient algorithm for testing the \\ 
	generic rigidity of graphs in the plane}
\author{C.~Moukarzel\footnote{Present Address: Instituto de F\'\i sica,
Universidade Federal Fluminense, \\ Niteroi RJ,
Brazil.\\email: cristian@if.uff.br} }
\address{
H\"ochstleistungsrechenzentrum, Forschungszentrum J\"ulich,\\ D-52425
J\"ulich, Germany.}
\maketitle
\begin{abstract}
Given a structure made up of $n$ sites connected by $b$ bars, the
problem of recognizing which subsets of sites form rigid units is not a
trivial one, because of the non-local character of rigidity in
central-force systems. Even though this is a very old problem of
statics, no simple algorithms are available for it so the most usual
approach has been to solve the elastic equations, which is very
time-consuming for large systems. Recently an integer algorithm was
proposed for this problem in two dimensions, using matching methods
from graph theory and Laman's theorem for two-dimensional graphs. The
method is relatively simple, but its time complexity grows as $n^2$ in
the worst case, and almost as fast on practical cases, so that an
improvement is highly desirable. I describe here a further elaboration
of that procedure, which relies upon the description of the system as a
collection of rigid \emph{bodies} connected by bars, instead of sites
connected by bars. Sets of rigidly connected objects are replaced by a
unique body, and this is done recursively as more rigid connections
between bodies are discovered at larger scales. As a consequence of
this ``rescaling transformation'', our algorithm has much improved
average behavior, even when its worst-case complexity remains $n^2$.
The time complexity of the body-bar algorithm is found to scale as
$n^{1.12}$ for the randomly diluted triangular lattice, while the
original site-bar version scales as $n^{1.9}$ on the same problem.
\end{abstract}
\section{ Introduction }
\label{sec:intro}
\begin{figure}[] 
\vbox{
\centerline{\psfig{figure=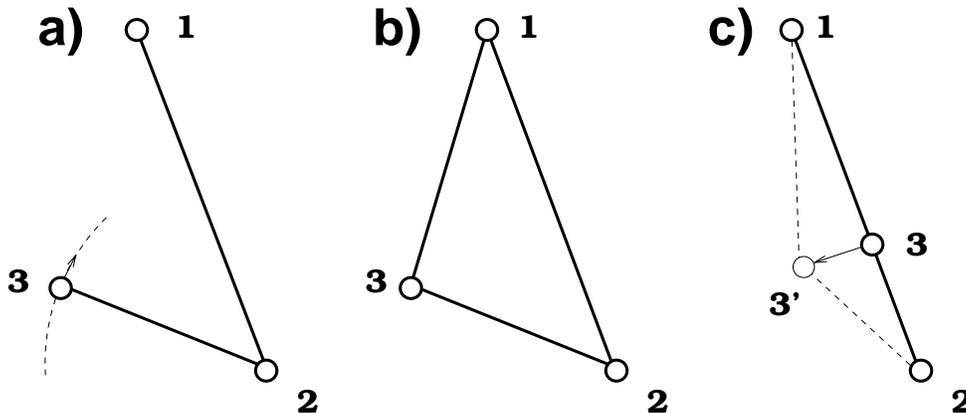,width=13cm,angle=270}}
\caption{ A simple illustration of the reasons by which a structure may
fail to be infinitesimally rigid (cases a) and c) ) is provided by a
triangle. Case a) is flexible since it admits a continuous deformation.
Case b) is both rigid (no continuous deformation is possible) and
infinitesimally rigid (no infinitesimal deformation is possible). Case
c) on the other hand provides an example of a \emph{degenerate}
configuration, for which the structure is rigid (there is no continuous
deformation leaving all bar lenghts unchanged) but not infinitesimally
rigid, since site $3$ may be desplaced by an infinitesimally small
amount, and all bar lengths would remain unchanged to first order in the
displacement. Failure to be infinitesimally rigid can be shown to be
equivalent to the existence of zero-frequency modes in the linear
approximation of the vibratory behavior of the structure. }
}
\label{fig:Gtriangle}
\end{figure}

Consider a structure~\cite{Lattice} made of $n$ sites connected by $b$
bars in $d$ dimensions. Determining the rigid properties of such
systems constitutes a problem of obvious technological interest, which
has been under study for a long time already. The first formal results
date back to Maxwell\cite{Maxwell}, who discussed the connections
between statics and geometry.  Posterior theoretical work in the
field~\cite{Laman,AsimovRoth,Crapo,Whiteley,TayWhiteley,Crapo2,TW,W2,W3},
has been accomplished by mathematicians and not widely known among
physicists, although the concept of rigidity appears in many fields of
physics as for example glasses~\cite{Thorpe,Franzblau2},  critical
phenomena~\cite{Percolation} and granular materials~\cite{Guyon} among
others.
\\
The physicist's treatment of this problem often reduced to a brute-force
solution of the elastic equations, because of the lack of a simple
integer algorithm for the identification of the rigid clusters. In this
work, recent progress in the design of such algorithms for the analysis
of rigid properties of generic lattices is reported.  To do so, let us
start by briefly introducing some of the basics of rigidity. In this
section the main ideas will we qualitatively described, leaving for next
section more precise definitions, which will be done with the aid of the
rigidity matrix. The interested reader is referred to the recent
literature for more complete
descriptions~\cite{Crapo,Whiteley,TayWhiteley}, alternative
approaches~\cite{Crapo2,GaWe} and recent results~\cite{TW,W2,W3} in the
field of rigidity. 
\\
A structure\cite{Lattice} is \emph{flexible} if it admits a continuous
deformation (a finite \emph{flexing}) preserving all bar lenghts, other
than the trivial translations and rotations in euclidean space.
Otherwise it is \emph{rigid}. Obviosuly if a structure admits a finite
flexing, then it also admits an infinitesimal flexing. If a structure
admits no non-trivial infinitesimal flexing, it is said to be
\emph{infinitesimally rigid}. Obviously a flexible structure is also
infinitesimally flexible. The converse is not always true, since there
may be special situations in which a structure is infinitesimally
flexible but does not have finite flexings. Figure~\ref{fig:Gtriangle}a
shows an example of a structure that admits a finite flexing, and
therefore is not rigid. If one more bar is added, the extra degree of
freedom corresponding to the flexing may be eliminated.
Figure~\ref{fig:Gtriangle}b is an example of a structure which is both
rigid and infinitesimally rigid. But the same triangle has special
combinations of site locations for which infinitesimal rigidity is lost.
This is exemplified in Figure~\ref{fig:Gtriangle}b, in which the three
sites are aligned. As a consequence  site $3$ may be displaced by a
small amount in the direction of the normal to bar $12$, and
all bar lenghts remain unchanged \emph{to first order in the
displacement}. Therefore this structure is infinitesimally
flexible though it is rigid to second order in the displacement.
\\
These situations for which rigidity does not imply infinitesimal
rigidity are very rare. They only occurr for a 'small' set of site
locations, which are called \emph{degenerate} configurations. The
complement of this set, the \emph{generic} configurations, form an open
dense subset of space. For a generic configuration, infinitesimal
rigidity is equivalent to rigidity. Configurations in which all site
coordinates are randomly chosen are with probability one generic (a
precise definition will be provided in next section).
The importance of infinitesimal rigidity can be easily understood in
physical terms. One can define a structure to be \emph{statically rigid}
if it is able to compensate, by means of suitable finite stresses in its
bars, any equilibrated load applied to its nodes. It is a classical
ingeneering result that static rigidity is equivalent to infinitesimal
rigidity (see \cite{TayWhiteley}) for a proof.). Therefore infinitesimal
rigidity, and not rigidity, is the relevant concept in the linear
approximation of elasticity. Case a) in Fig.~\ref{fig:Gtriangle} is
obviously not statically rigid. Case c) is also not statically rigid
because a load normal to bar $12$ applied on site $3$ cannot be
compensated by finite stresses on the bars, which are all parallel to
$12$.
\\
When studying the oscillatory properties of a structure in the harmonic
oscillator approximation, infinitesimal flexibility is equivalent to the
existence of degenerate, zero-frequency modes. Note that cases a) and c)
in Figure~\ref{fig:Gtriangle} both have zero-frequency modes, but for
different reasons. In case a) because there is no restoring force, and
in case c) because the restoring force is zero in the linear
approximation.  
\\
Throughout this work we will be concerned with the property of
infinitesimal rigidity, and therefore we drop the prefix in the
following.  As suggested by the example in Fig.~\ref{fig:Gtriangle},
a structure can fail to be rigid for two different reasons. In the first
place, because it has too few bars, or they are not correctly
distributed.  This has to do with the topology of the lattice, that is,
depends on its connectivity only, and not on the geometry of the
structure.  We will say that a given structure is \emph{generically
rigid} if it has the required minimum number of ``correctly
distributed'' bars. Generic rigidity is a necessary condition for
rigidity, but not a sufficient one.  The importance of generic rigidity
resides in the fact that a generically rigid structure will be rigid for
generic site locations. Generic rigidity  depends only on the
number and location of the bars, and not on site locations. In other
words, generic rigidity depends only on the topological properties of
the structure.
\\
Topological information about a structure is conveniently represented by
means of a graph $G=(V,E)$. Each site of the lattice is associated to a
node $a \in V$ , while bars are associated to edges $ab \in E$.  Nodes
$a$ and $b$ are \emph{adjacent} if edge $ab$ exists. Edge $ab$ is said
to be \emph{incident} to nodes $a$ and $b$ and, conversely, nodes $a$
and $b$ are incident to edge $ab$. Graphs as described here contain no
information about the geometry of the system (site locations), but only
about its connectivity properties. 
\\
A graph contains enough information about a structure if we are only
willing to discuss its generic properties, i.e. those which are valid
for ``almost all'' sets of site locations, except for  those few
degenerate configurations, which will be ignored.  It is only meaningful
to do this if degenerate configurations, and therefore lattices which
are generically rigid but not rigid, are ``exceptional''. A context in
which this is justified is when site locations (an assignment of site
locations to the nodes of a graph is called a \emph{realization} of the
graph) are randomly chosen.  This ensures that degenerate configurations
have zero probability to appear, i.e.  degenerate realizations are a
zero measure set. In view of this we would like  to determine, from
topological information only, whether a given graph is generically
rigid.  A first condition that must be satisfied is that the number $b$
of bars be large enough. Since $n$ points in $d$ dimensions have $dn$
degrees of freedom and each bar restricts one degree of freedom, $b$ has
to be at least equal to $dn - d(d+1)/2$\cite{n>d}, where $d(d+1)/2$ is the number
of distance-preserving linear transformations in $d$ dimensions ( $d$
translations and $d(d-1)/2$ rotations). But a right global count of bars
is not enough to ensure rigidity, since bars could be ``crowded'' on
certain subsets of the graph, while others have less bars than needed to
make them rigid. If a certain subgraph has more bars than necessary,
some of them are \emph{redundant}, and this subgraph is
\emph{overconstrained}. Bars which are not redundant are said to be
\emph{(generically) independent}. A sufficient condition for (generic)
rigidity is that the graph possesses $dn - d(d+1)/2$ (generically)
independent bars.  We see that the key point is being able to identify
independent bars. The basic theoretical tool for doing this in 2
dimensions is provided by a theorem due to Laman~\cite{Laman}.
\begin{laman} [Laman] \label{th:Laman}
The edges of a bar and joint graph $G=(V,E)$ are generically independent
in two dimensions if and only if no subgraph $G'=(V',E')$ has more than
2n'-3 edges.
\end{laman}
Laman's theorem constituted the first graph-theoretic characterization
of rigidity in 2 dimensions, but in its original form it would give a
very bad algorithm since it requires testing all possible subgraphs, of
which there is an exponentially large number. There are some equivalent
restatements of this theorem~\cite{Sugihara,TayWhiteley}, some of which
give rise to polynomial-time algorithms.  Using one such equivalence
due to Sugihara, Hendrickson~\cite{Hendrickson} has recently proposed
an algorithm for testing generic rigidity of two-dimensional graphs,
which is simple enough to admit an on-lattice
implementation~\cite{JacobsThorpe}.  Roughly described, Hendrickson's
algorithm consists in adding edges one by one to the graph and
matching~\cite{Matching1,Matching2} them to the nodes. If the matching
succeeds, the new edge is independent and is left on the system. If the
matching fails, then a) the set of edges visited during the failed
search is mutually rigid, and b) the last edge is redundant.  This
algorithm has a worst-case time complexity that scales as $O(n^2)$.
One factor of $n$ arises as edges are added one at a time, while the
degree to which the computational time is greater than $n$  is
determined by the typical size of the search that must be performed in
order to match each added edge.  If $O(n)$ sites  are mutually rigid,
this size is of order $n$, and the algorithm is of order $n^2$.
However in this case there is considerable time spent in searching over
edges that have been previously identified as mutually rigid.  There is
thus room for improvement, and this work is devoted to the description
of such an improved algorithm.
\begin{figure}[] \vbox{ 
\hbox{ {\bf a)}
\psfig{figure=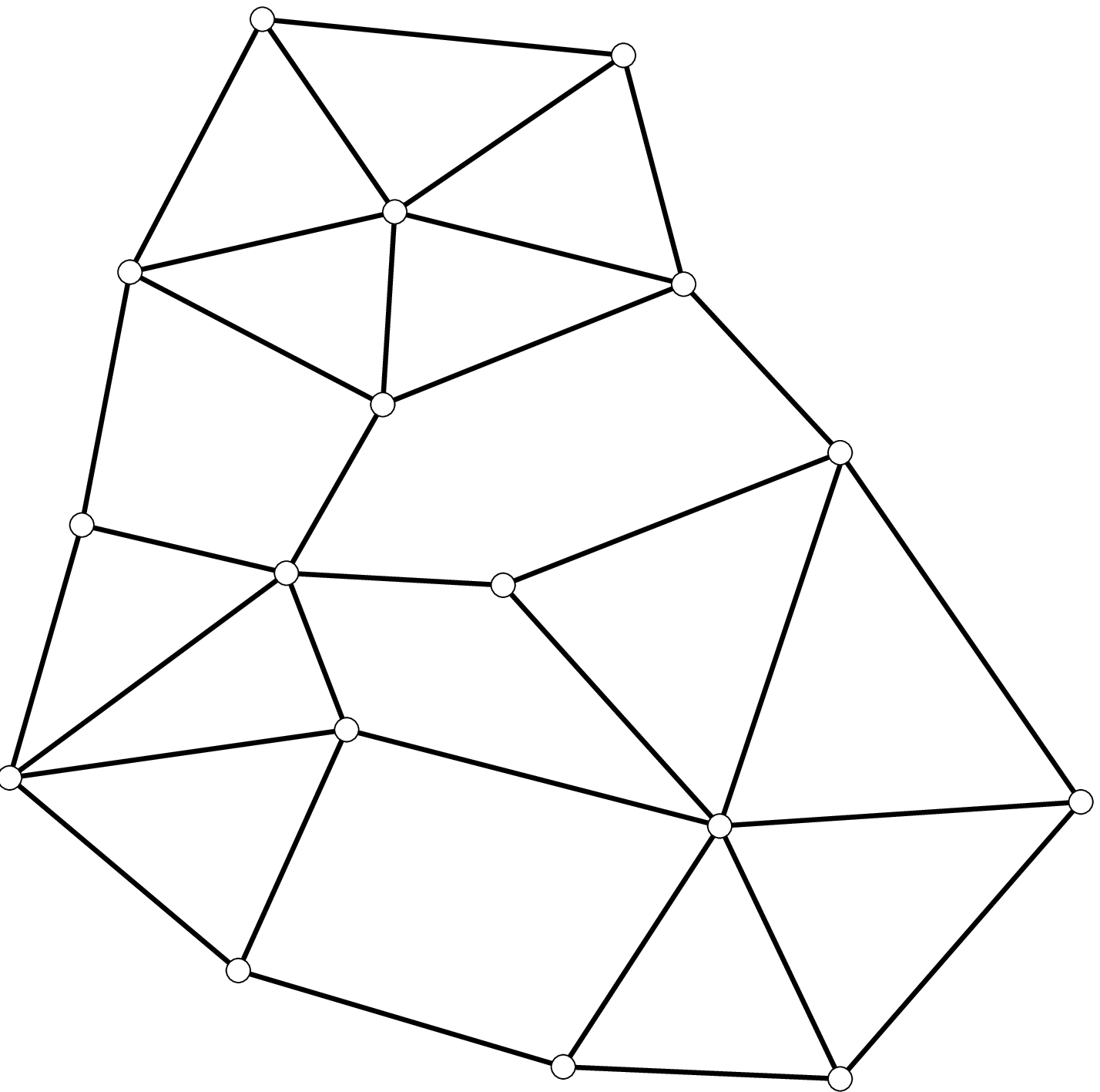,width=4cm} \hfill  {\bf b)}
\psfig{figure=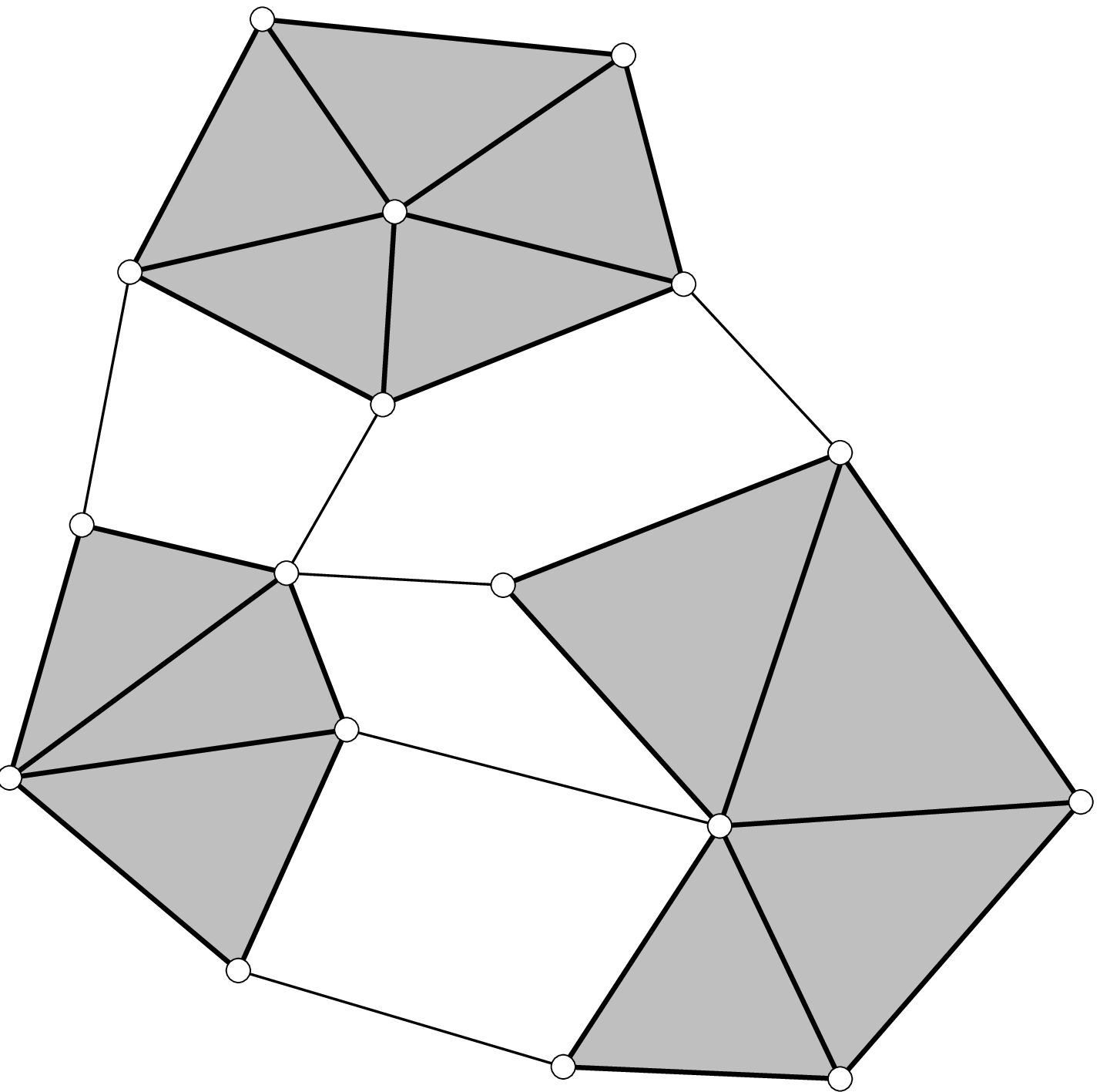,width=4cm}  \hfill {\bf c)}
\psfig{figure=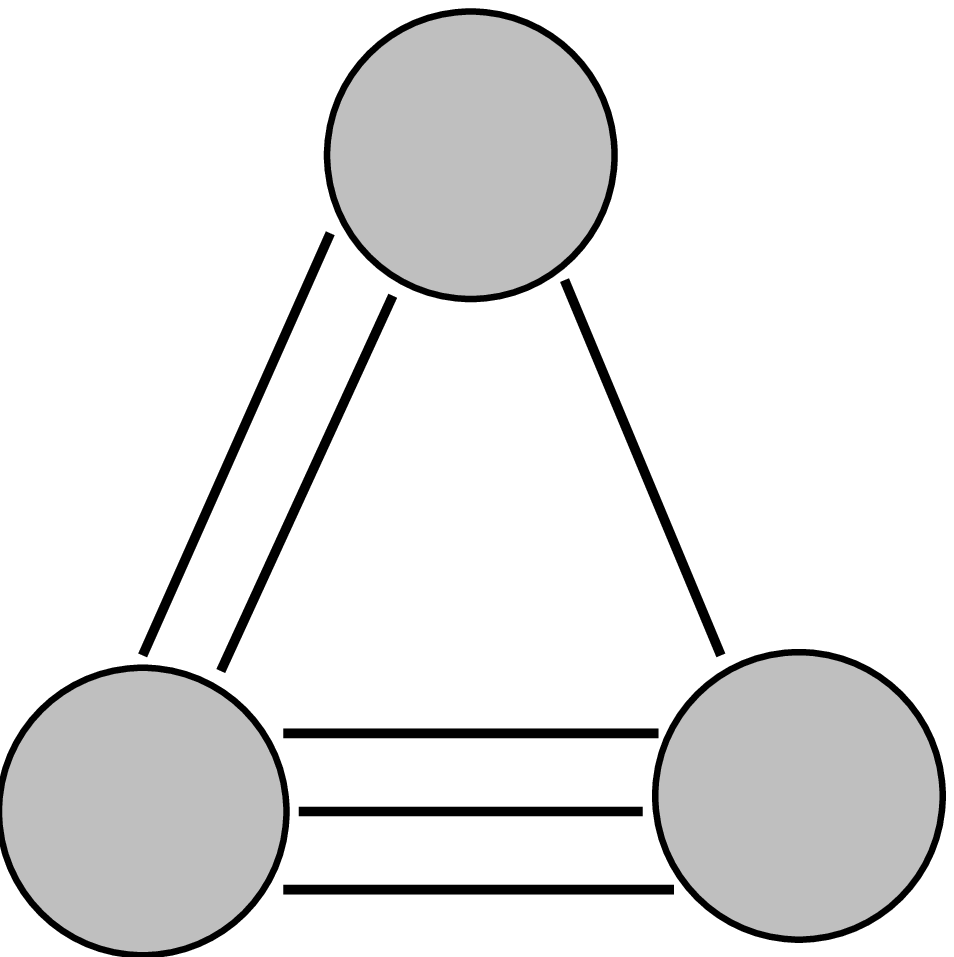,width=4cm}
}
\caption{ In the \emph{bar-joint} representation (a), each site of the
lattice is associated to a node of the graph, and bars are represented
by edges. If sub-sets of rigidly connected sites (b) are combined into
\emph{bodies} and represented by a node, we get the \emph{body-bar}
representation (c).  The resultant structure is a \emph{multigraph}
since several edges can connect a given pair of nodes. The process of
replacing a set of rigidly connected nodes by one node is called {\it
condensation}. }
\label{fig:bodies}
} \end{figure}
To avoid the need to search over edges which have already been
identified as mutually rigid, it is natural to combine mutually rigid
edges into clusters or
``bodies''~\cite{Tay,Guyon,Franzblau2,GaWe,W2,W3}.  To develop this idea
into a workable algorithm, we note that any bar and joint graph can be
considered as being composed of bodies and bars, where eventually some
of the bodies may be ``trivial'' bodies with just one site.  Each body
is represented by a node in a {\it multigraph}. An example of this is
shown in Fig.~\ref{fig:bodies}.  Each body, composed of several sites
rigidly connected by bars, is shown as a  node in the multigraph of
Fig.~\ref{fig:bodies}c. One of the advantages of the body-bar
representation is that the number of elements in the graph is smaller,
and may be reduced each time a cluster of rigidly connected bodies is
identified, by replacing them by just one node in a process we will call {\it
condensation}. 
\\
Of course we have now to demonstrate that the idea is sound, and for
that we must restate the relevant theoretical results~\cite{Hendrickson}
in terms of bodies and bars. We start in section~\ref{sec:rigmat} by
introducing the {\it rigidity matrix}, which will help us to more
precisely define the concepts of infinitesimal rigidity, generic
rigidity, generic configurations and of redundant and independent bars,
and also to discuss the necessary conditions for generic rigidity. In
section~\ref{sec:laman} we express Laman's theorem in the body-bar
language, while section~\ref{sec:algorithm} shows how Hendrickson's
algorithm is generalized for this case. Section \ref{sec:implementation}
discusses some practical implementation details. Also in this section
the performance of the body-bar algorithm here introduced is compared to
that of the previously existing joint-bar version. We will see that the
body-bar algorithm has much better scaling properties for the two
examples analyzed, even when its worst-case behavior is the same as that
of the joint-bar algorithm, that is, $O(n^2)$.
\\
\section{ Rigidity matrix for two-dimensional body-bar systems }
\label{sec:rigmat}
We will in the following only consider two-dimensional structures,
unless explicitly stated otherwise.  Consider Fig.~\ref{fig:bodies},
in which the relationship between a bar-joint and a bar-body
representation is explicitly drawn. Any subset of rigidly connected
bars and sites can be replaced by a \emph{body}, a rigid object that in
two dimensions has 3 degrees of freedom, two displacements and one
rotation. We call all remaining bars \emph{external}. Each body has a
set of joints on its surface, to which external bars are incident. We
let $x_i$ be the location of joint $i$ belonging to body $a$.  An
infinitesimal motion is a set of instantaneous velocities $\{v_i\}$,
one for each joint, which leave all bar lengths unaltered. This 
condition is written:
\begin{equation}
\label{eq:condition}
(x_i -x_j)  \cdot  (v_i - v_j) = 0
\end{equation}
for every bar $ij$ with $i \in a, j \in b$. We now express the
velocities of the joints in terms of the velocities of the body to
which they belong.  For this we select an arbitrary point $x_a$ for
each body $a$, and say that the velocity $v_i$ of any joint $i$ of a
body is equal to the velocity $v_a$ of $x_a$ plus a rotational
component, which is $\omega_a \wedge (x_i-x_a)$, where $\omega_a$ is
the angular velocity (a vector normal to the plane) of body $a$, and
$\wedge$ indicates vector product.
\begin{equation}
v_i = v_a + \omega_a \wedge (x_i - x_a)
\end{equation}
Without loss of generality, we can choose all those arbitrary reference
points $x_a$ to be at the origin and get,
\begin{equation}
\label{eq:v}
v_i = v_a + \omega_a \wedge x_i 
\end{equation}
Now let us rewrite  (\ref{eq:condition})  by using (\ref{eq:v}), so
that
\begin{equation}
\label{eq:rewritten}
( x_i - x_j ) \cdot \{ (v_a-v_b) + \omega_a  \wedge x_i 
- \omega_b \wedge x_j \} = 0
\end{equation}
A little algebra shows that (\ref{eq:rewritten}) can be reduced to,
\begin{equation}
\label{eq:final}
(x_i-x_j) \cdot (v_a-v_b) + (x_j \wedge x_i) \cdot (\omega_a-\omega_b) = 0
\end{equation}
This set of equations can be formally written as
\begin{equation}
\label{eq:matform}
\tilde M \vec V = 0
\end{equation}
where $\vec V$ is the vector of velocities and contains 3 entries for
each body. Each row in $\tilde M$ corresponds to a bar $ij$. Conditions
(\ref{eq:condition}), when written in the bar-joint representation, also
give rise to a matrix equation of the form (\ref{eq:matform}). In the
bar-joint representation only the first term in (\ref{eq:final}) occurs,
each row of the rigidity matrix is associated with the vector
$(x_j-x_i)$, and there are 4 non-zero elements per row.  In the body-bar
case, each row of the rigidity matrix is associated with the `line-bound vector''
$(x_j-x_i, x_i \wedge x_j)$  and there
are 6 non-zero elements per row since $x_i \wedge x_j$ is a pseudoscalar
in two dimensions.  Line-bound vectors represent a force acting along a
line~\cite{Crapo} and, in contrast to vectors, are not translationally
invariant. They are only invariant under translations in the direction
($x_i-x_j$), meaning that a force can be moved along its line of action
without changing its effect. Line-bound vectors have three independent
components in 2d, two of them are needed to determine the vector and the
third one locates its line of action in the plane.
\\
We will consider the general case of multigraphs formed by $n$ bodies
and $m$ point-like (or ``trivial'') bodies, as in
Fig.~\ref{fig:generic}, which we denote as $G(n,m)$.    The general
equations (\ref{eq:final}) or (\ref{eq:matform})   hold for $G(n,m)$
with the additional constraint that the angular velocity $w_a=0$ for
each of the $m$ point-like bodies.  These additional constraints reflect
the fact that rotation is irrelevant for them, so their angular
velocities can be arbitrarily fixed, thus reducing the dimension of
$\vec V$.  Counting the number of degrees of freedom, we then find (3n +
2m).  The space of solutions of (\ref{eq:matform}) has at least
dimension 3, since two rigid translations and a rotation of the system
as a whole leave all bar lengths unchanged.  This means that the rank of
$\tilde M$ cannot be larger than $K(n,m) = 3n+2m-3$.  The system is said
to be (infinitesimally) rigid if the rigidity matrix has this maximal
rank $K(n,m)$, which means that the \emph{only} infinitesimal motions
are the Euclidean rigid transformations.
\\
A \emph{realization} is an assignment of site locations to all nodes of
a graph. For certain (degenerate) realizations, the rank of the rigidity
matrix may be accidentally lowered by the existence of algebraic
dependencies between the node coordinates (degeneracies). This will only
happen if a determinant is zero, and given that determinants of the
rigidity matrix are polynomials in the site coordinates, degenerate
realizations must satisfy a finite number of polynomial equations.
Therefore the subset of configurations for which the rigidity matrix
attains the maximum possible rank over all sets of coordinates
constitutes an open dense subset. We say that a realization is
\emph{generic} if all site coordinates are algebraically independent
over the rationals. Therefore at a generic configuration the rank of the
rigidity matrix attains its maximum value over the site coordinates.
\begin{figure}[] 
\vbox{
\centerline{\psfig{figure=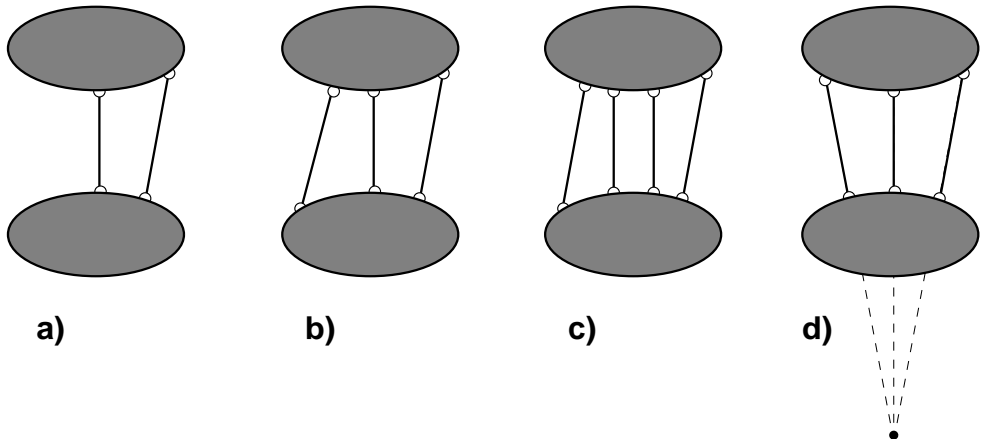,width=13cm,angle=270}}
\caption{ A simple body-bar structure for which: a) the bar set is
independent, and the structure is flexible; b) the bar set is
independent, and the structure is rigid; c) the bar set is dependent due
to an excess of bars, and the structure is rigid; and c) the bar set
becomes dependent at a degenerate configuration, and the structure
becomes infinitesimally flexible there. In this last case, the existence
of a common intersection point for three bars is a non-generic
configuration at which the rank of the associated rigidity matrix is
reduced from its generic value of 3 to a non-generic value 2.   The
extra eigenvector is identified as an infinitesimal relative rotation of
the two bodies around the common intersection point. }
}
\label{fig:Gbodies}
\end{figure}
A set of edges is \emph{independent} if their associated rows in $\tilde
M$ are linearly independent in an algebraic sense. This means that an
edge will be independent if and only if by removing it the rank of the
rigidity matrix is decreased by one. Cases a) and b) in
Fig.~\ref{fig:Gbodies} are independent, but case a) is flexible while
case b) is infinitesimally rigid.
\\
If the removal of a given edge $e$ does not alter the rank of $\tilde
M$, $e$ is \emph{dependent} or \emph{redundant}. Dependencies can arise
because there are too many bars (for example the edge set in
Fig.~\ref{fig:Gbodies}c) or because of degenerate configurations. (for
example three bars with a comon point as in Fig.~\ref{fig:Gbodies}d).
The difference is that case c) is infinitesimally rigid while case d) is
not. The dependency produced by the coincident intersection point has
reduced the rank of the rigidity matrix from its maximum value of $3$ to
a value of $2$. There is therefore one extra infinitesimal motion, which
can be identified as a relative rotation of the two bodies around the
intersection point of the bars.  A similar reasoning holds for the
example shown in Fig.~\ref{fig:Gtriangle}c.
\\
Once the atypical character of degeneracies is recognized, we are
justified in ignoring them, and concentrate on generic properties only.
Therefore we define generic rigidity by saying that: \emph{A structure
is generically rigidy if its rigidity matrix attains its maximum rank
$K(n,m)$ at a generic configuration.} The relevance of generic
properties  is ensured by a theorem due to
Gluck~\cite{Gluck}.
\begin{generic} [Gluck] \label{th:Gluck} 
If a graph has a single infinitesimally rigid realization, then all its
generic realizations are rigid.
\end{generic}
In other words, \\
\emph{At a generic realization, a structure is infinitesimaly rigid if and
only if it is higher-order rigid if and only if its multigraph is
generically rigid}
\\
In addition to the already defined property of generic site locations,
we will assume the following \emph{generic incidence condition} to hold:
{\it no 3 bars can be incident to the same joint of a \emph{non-trivial}
body}. This condition means that  bars incident to the same body are
located on generic lines. In the following,  whenever we refer to
generic realizations of multigraphs we will be assuming generic joint
locations as well as generic incidence. Notice that multiple incidences
 are allowed if they occur on trivial (single-site) bodies. An example
of a multigraph that is generic in the sense required here is shown in
Fig.~\ref{fig:generic}.
\vbox{ 
\begin{figure}[] 
\centerline{ \psfig{figure=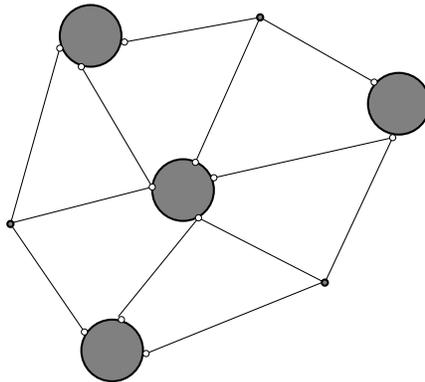,height=5cm} }
\caption{ A multigraph that satisfies the generic incidence condition.
No surface joint has more than two incident bars. Point-like bodies can
have an arbitrary number of incident bars.}
\label{fig:generic}
\end{figure} 
	}
Throughout this work we will assume all multigraphs $G(n,m)$ to contain at
least one non-trivial body or two point-like ones, that is, $n+m/2 \geq
1$. Excluded are then the graph with no nodes and those with just one
trivial (point-like) body. Under this condition the following results
follow almost trivially from our discussion of the rigidity matrix.
\begin{generic} \label{th:hask}
Every rigid multigraph $G(n,m)$  has a subset of $K(n,m)$ independent bars.
\end{generic}
\begin{generic} \label{th:ifmore}
If a multigraph $G(n,m)$ has more than $K(n,m)$ bars, some are dependent.
\end{generic}
\begin{generic} \label{th:ifrigid}
If a multigraph $G(n,m)$ with exactly $K(n,m)=3n+2m-3$ bars
is rigid, then there is no subgraph $G'$ with more than $K(n',m')$ bars.
\end{generic}
 
In the process of obtaining the bar-body representation from the
bar-joint representation, we have replaced subsets of rigidly connected
sites by bodies with 3 degrees of freedom. This does not change the
number of independent infinitesimal motions of the system, since these
subsets, being rigid, had 3 degrees of freedom each. This means that
the dimension of the null space of the rigidity matrix is the same in
the bar-joint and bar-body representations. A consequence of this is
the following.
\begin{generic} \label{th:alwaysdep} 
A set $E$ of external edges is dependent in the body-bar representation
if and only if it is dependent in the corresponding joint-bar
representation.
\end{generic}
{\bf Proof.} 
The removal of an independent edge causes the rank of the rigidity
matrix to change, increasing by one the number of independent
infinitesimal motions. Assuming a given edge $e \in E $ to be dependent
in one representation but not in the other would then conflict with our
discussion above.$\hfill~\diamond $ 
\\
Another result which we need for  later use is the following. 
\begin{generic} \label{th:four}
If an edge $e$ incident to a non-trivial(trivial) body $a$ and to some
other body $b \neq a$ is dependent, then there exist at least 3(2)
other edges $\{e',e'',e'''\} (\{e',e''\})$ incident to $a$ which are
also dependent.
 \end{generic}
{\bf Proof:}  Let $a$ be a non-trivial body and assume that and edge $e$
incident to $a$ is dependent. Take the row of $\tilde M$ corresponding
to $e$, which under the hypothesis can be expressed as a linear
combination of other rows of $\tilde M$, and consider its 3 components
associated to the 3 degrees of freedom of $a$.  These are the 3
components of a line-bound vector, and are independent degrees of
freedom in a generic realization so that at least 3 other line-bound
vectors (rows of $\tilde M$) are needed to express it as a linear
combination. The demonstration for the case in which $a$ is a trivial
body is similar, except that vectors (2 degrees of freedom) take the
place of line-bound vectors, and therefore only 2 other edges are
needed.  $\hfill~\diamond $ \\
We have introduced the rigidity matrix $\tilde M$ through a discussion
of infinitesimal rigidity, but it also appears in the context of static
rigidity, describing how external loads are resolved into stresses on
bars. Dependent subgraphs can then be identified to be those that can
sustain equilibrated internal stresses even in the absence of external
loads.
\\
\section{Laman's theorem for body-bar systems}
\label{sec:laman}
A general graph-theoretic characterization of rigidity for linkages of
bodies in $n$-space was first provided by Tay~\cite{Tay}. We will
derive a similar result in two dimensions, which holds for the case of
mixed multigraphs, and which we need for our algorithm. Mixed
multigraphs are those  including  point-like bodies as well as
non-trivial ones. Our derivation is based on Laman's theorem for
bar-joint systems.
\begin{figure}[] \vbox{ 
\centerline{ \psfig{figure=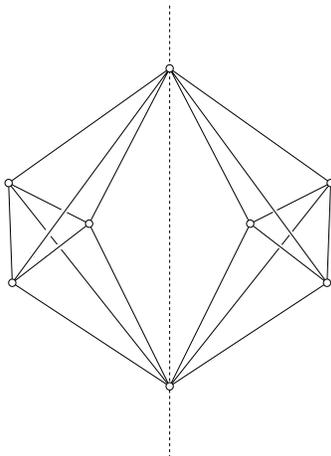,height=6cm} }
\caption{ The graph in this figure has no rigid realization in 3
dimensions, even when it has all the required $18 = 3 \times 8 -6$ bars,
and no subgraph of it violates Laman's condition $b' \le 3 n' -6$. }
\label{fig:counterexample}
} \end{figure}
Theorems \ref{th:hask} and \ref{th:ifrigid} imply that a rigid system
must have a set of $K(n,m)$  well distributed bars, where the meaning of
``well distributed'' is that no subgraph has ``too many'' bars. This
condition is known, in the context of bar-joint systems, as Laman's
condition, and is a \emph{necessary} condition for rigidity in
\emph{any} space dimension (in a suitably generalized form). The
converse, i.e. $K(n,m)$ well distributed bars $\Rightarrow$ rigid does
not hold in any dimension above 2. A counterexample in 3 dimensions, due
to Whiteley, is shown in Fig.~\ref{fig:counterexample}. This structure
satisfies $b' \leq 3 n' - 6$ for all subgraphs with $n' > 2$, yet it is 
dependent and therefore non-rigid.
\\
Laman was able to show~\cite{Laman} that the ``correct distribution'' of
bars is a sufficient condition for rigidity in 2 dimensions (Theorem
\ref{th:Laman} above). Here we translate his result to the body-bar case
and show that
\begin{generic} 
\label{th:extended}
The edges of a multigraph $G(n,m)=G(V_n,V_m,E)$ are independent in 2d if
and only if there is no submultigraph $\tilde G$ with more than \mbox{$3
\tilde n + 2 \tilde m -  3$} edges.
\end{generic}
{\bf Proof.}  The demonstration of necessity, i.e. independent
$\Rightarrow$ well distributed, follows trivially from Theorem
\ref{th:ifmore} above.  In order to demonstrate sufficiency, we will
transform the body-bar graph to a bar-joint graph and use Laman's
theorem to show that, if a multigraph is dependent, there is a subset of
it with too many bars, i.e. a bad submultigraph.
\\
Assume that a multigraph $G=(V_n,V_m,E)$ contains a subset $E'$ of
dependent bars. Let $G'=(V'_n,V'_m,E')$ be the submultigraph of $G$
defined by restricting the edge set to $E'$ and the node sets to those
nodes incident to some edge in $E'$. For each non-trivial body $a \in
V'_n$, let the cardinality $g_a$ be the number of joints on its surface
to which bars $ab \in E'$ are incident. According to
Theorem~\ref{th:four}, in order for a non-trivial body $a$ to belong to
$V'_n$, at least 4 of its incident bars must belong to $E'$.  This
together with the condition of generic incidence
(Section~\ref{sec:rigmat}) implies \mbox{$g_a \ge 2$.}  A body of
cardinality \mbox{$g_a \ge 2$} can be replaced  by an isostatic
bar-joint graph $G_a$ made up of \mbox{$2g_a - 3$} well distributed
``internal'' bars and $g_a$ point-like bodies, these last taking the
place of  surface joints.  Doing this for each of the $n'$ non-trivial
bodies in $G'$ leads to the \emph{expanded} graph $G_E$.

Theorem~\ref{th:alwaysdep} implies that $E'$ is also dependent in the
expanded graph $G_E$, and since Laman's theorem in its original form
applies to it, there must be a subgraph $\tilde G_E=(\tilde V,\tilde E)$
of $G_E$ with $\tilde b$ bars and $\tilde m$ joints such that \mbox{$\tilde b
> 2 \tilde m -3$.}

\begin{figure}[] \vbox{ 
\centerline{\psfig{figure=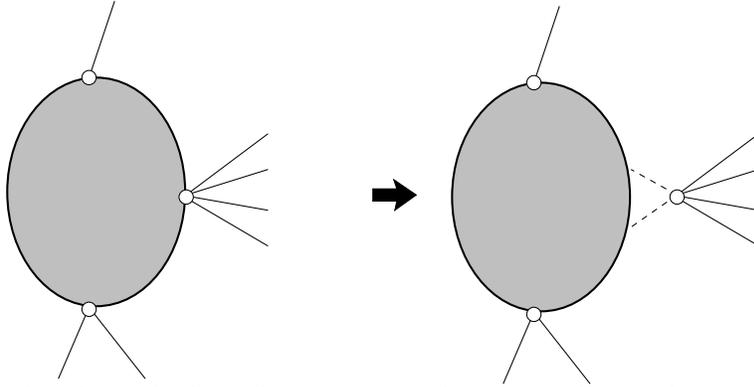,width=10cm} }
\caption{ A graph with a multiple-incidence joint (a surface joint with
more than 2 incident bars) is not generic. An equivalent (with the same
rigid properties) graph that is generic can be obtained by detaching the
surface joint and connecting it to the body with two auxiliary
bars (dashed lines). }
\label{fig:auxiliary}
} \end{figure}

We will have in general $\tilde b = \tilde b_e + \tilde b_i$, where
$\tilde b_e$ is the number of external edges and $\tilde b_i$ the number
of internal edges in $\tilde E$.
 $\tilde E$ cannot be entirely formed by internal edges since
they are well distributed, and therefore it must contain one or more
external edges. Two cases are possible: \\
\noindent
{\bf a)}  There are no internal bars in $\tilde E$.
\\
If this is the case none of the nodes of $\tilde G_E$ can be the
surface joint of a body. The reason for this is Theorem~\ref{th:four}.
At least 3 bars incident to a point are dependent if one of them is. But
if none of them is internal the point cannot be a surface joint since at
most 2 incident bars are allowed per surface joint in generic
multigraphs.  This means that no bar in $\tilde E$ is incident to
a subgraph $G_a$. In this case the bad subgraph is entirely formed by
point-like bodies and the result follows immediately since  Laman's
theorem is a particular case of this one (with $n=0$).\\
\noindent
{\bf b)} There are one or more internal bars in $\tilde E$, or equivalently,
there is at least one edge $e \in \tilde E$ incident to $G_a$ for some body
$a$. \\
If this edge $e$ has both ends incident to the same $G_a$, then we have
identified the bad sub-multigraph as being formed by body $a$ and this
edge with both ends connected to it, since \mbox{$1=b>3 n-3=0$.}
Then assume that  no edge $e \in \tilde E$ has both ends connected to the
same $G_a$. For each body $a$ to which some edge $e \in \tilde E$ is
incident, let $\tilde G_a$ with $\tilde g_a$ nodes and $\tilde b_a$
edges, be the subgraph of  $G_a$  contained in $\tilde G_E$. Let $\tilde
n_e$ be the number of such bodies, and $\tilde m_e$ the number of joints
in $\tilde G_E$ other than surface joints. The condition that $\tilde
G_E$ be a bad subgraph is then rewritten
\begin{equation} \label{eq:bad}
\tilde b_e +  \sum^{\tilde n_e}_{a=1} \tilde b_a > 2 ( \tilde m_e +
\sum^{\tilde n_e}_{a=1} \tilde g_a) - 3
\end{equation}
Theorem~\ref{th:four}, together with the generic incidence condition,
ensure that $\tilde g_a \ge 2$. Since $G_a$ are independent by construction
Laman's condition then implies  $\tilde b_a \leq 2 \tilde g_a -3$ . Using this
and (\ref{eq:bad}) we get
\begin{equation} \label{eq:bad2}
\tilde b_e +  \sum^{\tilde n_e}_{a=1} (2 \tilde g_a -3 ) > 2 ( \tilde m_e
+   \sum^{\tilde n_e}_{a=1} \tilde g_a) - 3
\end{equation}
, or
\begin{equation} \label{eq:bad3}
\tilde b_e > 2  \tilde m_e +   3 \tilde n_e - 3
\end{equation}
which finishes the demonstration for  generic body-bar
graphs.~$\hfill~\diamond $ \\ Non-generic graphs, i.e. those with more
than 2 bars incident to a joint of a body (we call this ``multiple
incidence'') can also be handled by transforming them into an equivalent
generic graph in the following form. Notice that there is no limitation
on the number of bars incident to point-like bodies. Thus we can simply
replace each multiple-incidence joint of a body by an auxiliary
structure made of a point-like node connected to the body by two new
bars, like in Fig.~\ref{fig:auxiliary}. The graph obtained by
transforming in this way all multiple-incidence joints is  generic in
the sense required here, and therefore the extended Laman's theorem
(theorem~\ref{th:extended} above) applies to it. It is easy to see
that this transformed graph has equivalent rigid properties.
%
\section{The algorithm}
\label{sec:algorithm}
\begin{figure}[] \vbox{ 
\hbox{ {\bf a)} 
\psfig{figure=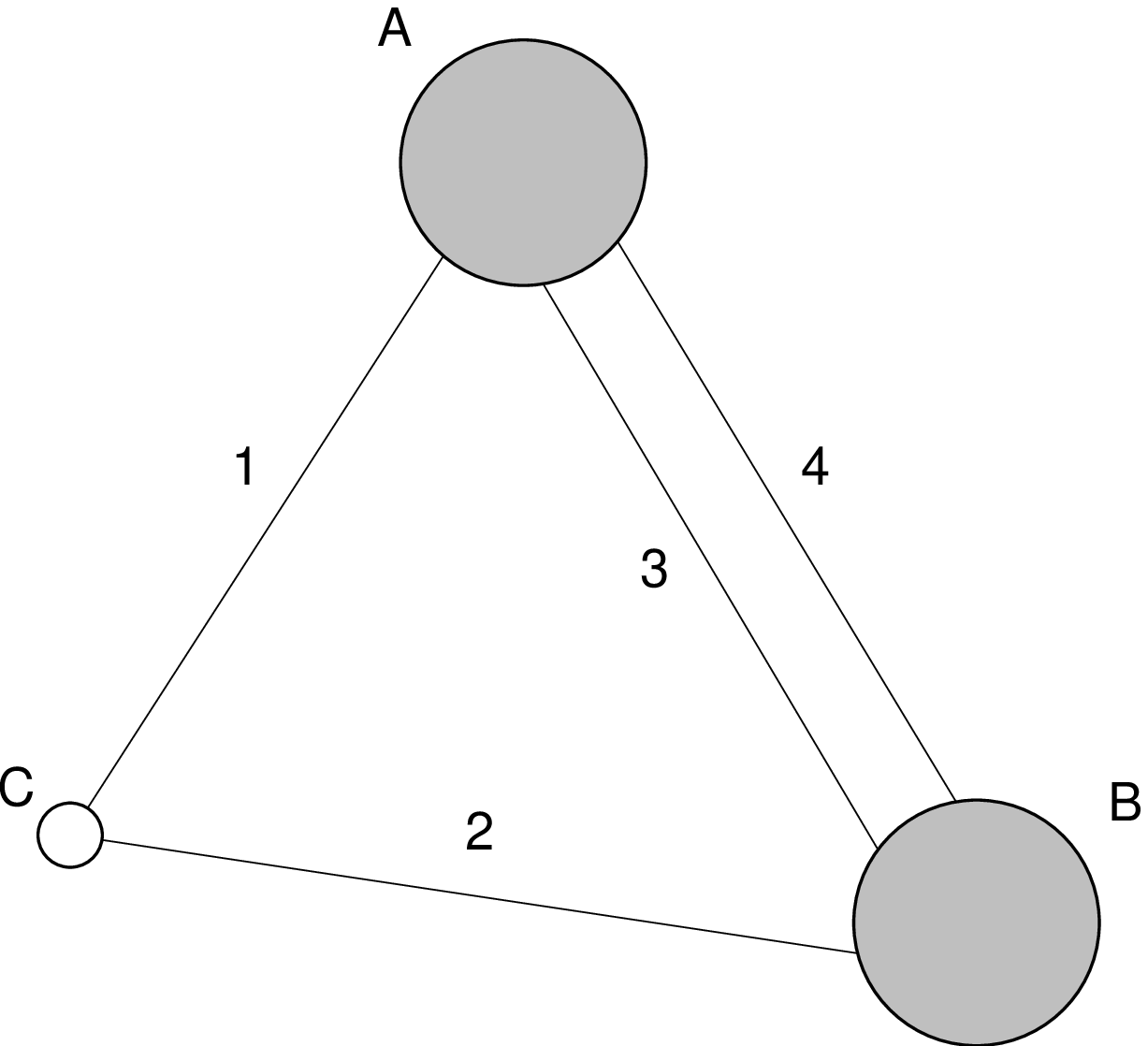,width=5cm} \hfill {\bf b)} 
\psfig{figure=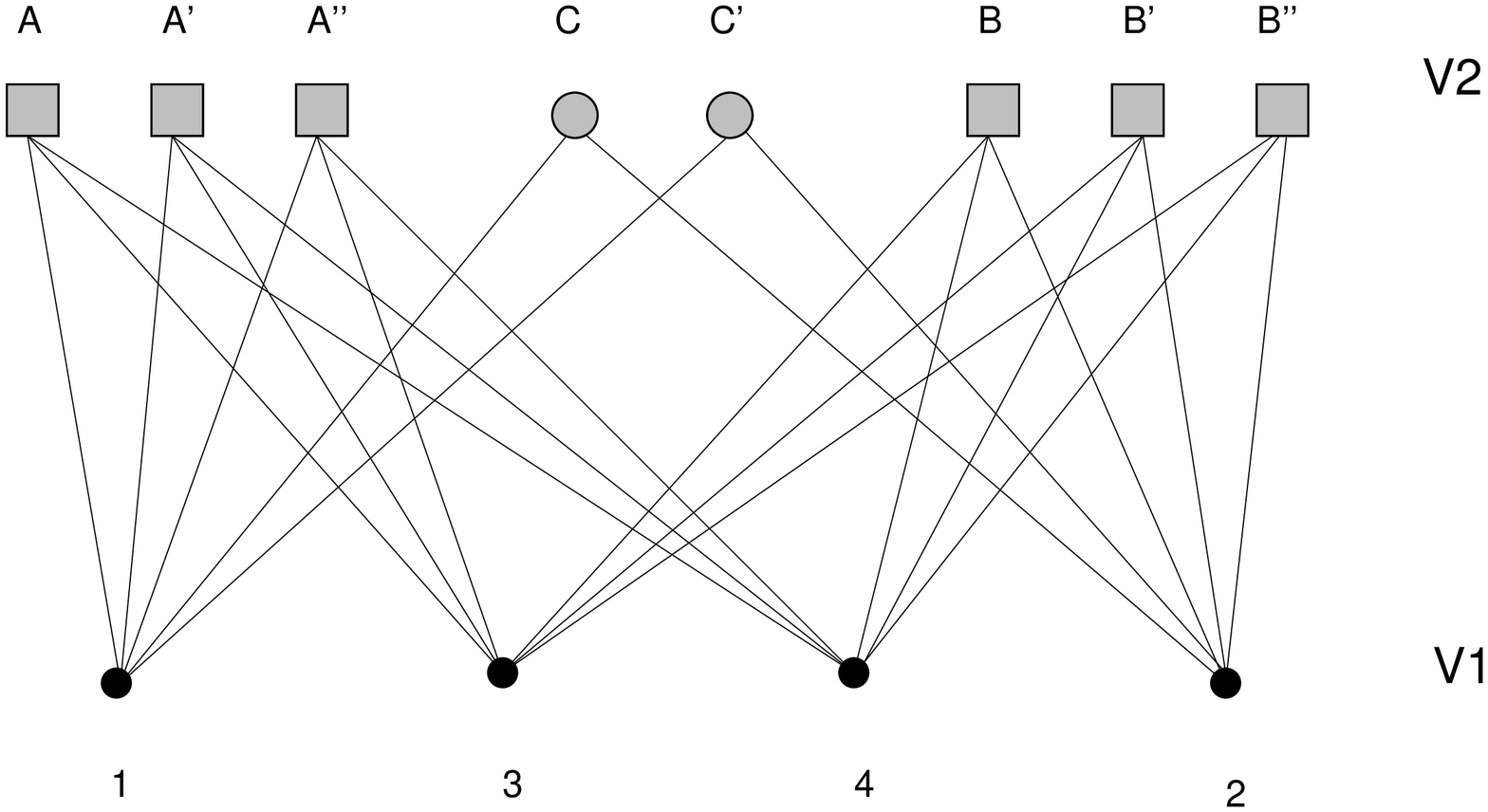,width=9cm} 
}
\caption{  The original graph $G(V_n,V_m,E)$ has $V_n = \{ A,B\}$
(non-trivial bodies), $V_m=\{C\}$ (point-like bodies) and
$E=\{1,2,3,4\}$ (edges). The bipartite graph $B(G)$ derived from $G$ has
as node set the union of $V1$ and $V2$. $V1$ is the edge set of $G$,
while $V2$ is made of 3 copies of $V_n$ plus 2 copies of $V_m$. Elements
of $V1$ and $V2$ are connected by and edge in $B(G)$ if they are
incident in $G$. }
\label{fig:bipartite}
} \end{figure}

\begin{figure}[] \vbox{ 
\centerline{\psfig{figure=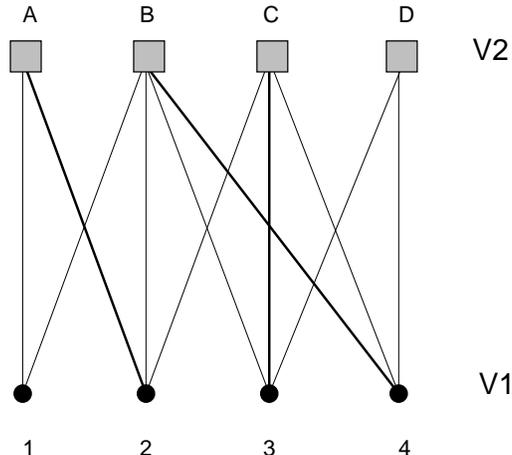,height=6cm} }
\caption{   A matching $\mathcal{M}$ (thick lines) can be enlarged to
cover node $1$ if an augmenting path is found. Starting from $1$ all
unmatched (thin) edges are followed to nodes in $V2$, and from them all
matched edges are followed back to nodes in $V1$, in a breadth-first
manner.  Repeating this procedure, an exposed node $D$ is found in
$V2$. If now thin and thick edges are interchanged along the path $1
\to D$, the enlarged matching is obtained.}
\label{fig:matching}
} \end{figure}

Laman's theorem constitutes a graph-theoretic characterization of
rigidity, but a naive implementation of it, namely checking all
possible subgraphs, would give a very poor algorithm.
Sugihara~\cite{Sugihara} and later Hendrickson~\cite{Hendrickson} have
used a reformulation of Laman's theorem to propose efficient (polynomial-time)
algorithms for this problem. This section follows  Hendrickson's
approach, adapting his arguments to the body-bar case where needed.
\\
We first define the bipartite graph $B(G)$ generated by a graph
$G(V_n,V_m,E)$ in the following way: $B(G)$ is composed of two sub-sets
of nodes $V1$ and $V2$, and a set of edges connecting nodes of $V1$
with nodes of $V2$. There are no edges between nodes in the same subset
(that is what defines a bipartite graph).  The first subset $V1$ is
the set of edges $E$ of $G$, while $V2$ is composed of 3 copies of the
set of non-trivial nodes $V_n$ plus 2 copies of the set of point-like
nodes $V_m$. Edges of $B(G)$ connect the edges $(ab)$ of $G$ ( $V1$
nodes) to all copies ($V2$ nodes) of the bodies $a$ and $b$ to which
$(ab)$ is incident. An example of this is shown in Fig.~\ref{fig:bipartite}.  
We now briefly describe some concepts from graph theory which we need
for our algorithm. The reader is referred to the literature on the
subject (\cite{Matching1},\cite{Matching2}) for detailed accounts. 
A \emph{matching} $\mathcal{M}$ of a graph is a subset of edges, no two
of which share a node.  Edges in $\mathcal{M}$ are said to be
\emph{matched}, and edges not in $\mathcal{M}$ \emph{unmatched}. Nodes
incident to a matched edge are \emph{covered}, while the rest are
\emph{exposed}. If $(ab) \in \mathcal{M}$, nodes $a$ and $b$ are
\emph{mates}. A matching $\mathcal{M}$ is \emph{maximum} if it has the
largest possible number of edges.  A matching is \emph{perfect} or
\emph{complete} if no node is exposed. Obviously, a perfect matching is
also maximum. The \emph{matching problem} (finding maximum matchings)
is a classic in graph theory, and has many practical applications. A
\emph{path} is a chain of edges of the form $\{(ab) (bc) (cd) \cdots
\}$, and it is \emph{alternating} if matched and unmatched edges follow
each other in the succession. An alternating path $\{ (ab) (bc) \cdots
(xy) \}$ is an \emph{augmenting path} if both $a$ and $y$ are exposed
nodes. The name is due to the obvious fact that any such path allows
one to increment the number of edges in the matching by one, simply by
interchanging \mbox{\it matched $\leftrightarrow$ unmatched } along the
path. Moreover, the problem of finding a maximum matching can be
reduced to that of finding augmenting paths, since
\begin{generic}   \label{th:noaugmenting}
$\mathcal{M}$ is  maximum $\iff$ there are no augmenting paths.
\end{generic}
Then a maximum matching is found by repeatedly discovering and
inverting augmenting paths.  A particularly simple case is the matching
of bipartite graphs, i.e. those whose node set $V$ can be partitioned
into two subsets $V1$ and $V2$ such that no edge is incident to two
nodes in the same subset. The search for augmenting paths can be
efficiently done by growing {\it Hungarian Trees} from exposed nodes.
They are built by breadth first search (BFS) in the following way.
Consider Fig.~\ref{fig:matching} in which thick lines represent edges
in the current matching. We start the search for augmenting paths from
an exposed node $v1 \in V1$. Take node $1$ for example. Following all
unmatched (thin) edges incident from $v1$, go to nodes $v2$ in $V2$.
Since $v1$ is exposed it is incident to unmatched edges only. In our
example of Fig.~\ref{fig:matching} these edges lead to $A, B$. If any
of these $v2$ nodes is exposed, we have found an augmenting path. If
not, their mates in $v1$ are sent to a \emph{queue} $Q$ for further
inspection. These mates are $2, 4$ in this case.  Nodes $v2$  are
marked \emph{visited}.  Once all neighbors of $1$ have been exhausted
without finding an exposed node, the next element $v1$ in queue $Q$ is
taken.  Say it is node $2$.  All unmatched edges incident to $2$ are
followed to $v2$ in $V2$ and, if some of them leads to an exposed node,
the search is over. Otherwise, and if $v2$ was not visited before, it
is marked visited and its mate in $V1$ is sent to $Q$. In our example
this would result in node $3$ being sent to $Q$.  The search
proceeds in this way until either an exposed node in $V2$ is found , or
$Q$ is depleted. In our example, after taking node $4$ from $Q$ we
would find node $D$ exposed. The matching  can then be enlarged to
cover node $1$ by inverting the path $D \to 4 \to B \to 1$. If on the
other hand  $Q$ is depleted without finding an exposed node in $V2$,
there is no augmenting path from $1$. We say in this case that
$v_1$ cannot be matched.
Now we are ready to discuss some results needed for our algorithm.
Using the relation between $G$ and $B(G)$ we can give an equivalent
formulation for the extended Laman's theorem, which will provide the
basis for our algorithm.
\begin{generic}
\label{th:equivalences}
The following are equivalent:\\
A) The edges of $G$ are independent in 2 dimensions.\\
B) For every edge $ab$ in $G$, the multigraph $G_{ab}$ formed by
quadrupling $ab$ has no subgraph with $b'>3n'+2m'$.\\
C) For each edge $ab$ in $G$, the bipartite graph $B(G_{ab})$ has no
subset of $V1$ that is adjacent only to a smaller subset of $V2$.\\
D) For each edge $ab$ in $G$, $B(G_{ab})$ has a complete matching from
$V1$ to $V2$.
\end{generic}
{\bf Proof:} The equivalence of A) and B) is a trivial consequence of
Laman's theorem in its extended form.  The equivalence of B) and C) is
an immediate consequence of the way in which $G_{ab}$ is constructed.
If such a subset exists, one would have $b'$ nodes of $V1$ connected
only to $3n'+2m'<b'$ nodes of $V2$.  To prove that C) and D) are
equivalent, it is easy to first see that C) is necessary for D).  To
prove that D) is necessary for C), assume that no complete matching
exists.  Then there exists at least one exposed node $x$ in $V1$.  Do a
breadth first search (BFS) in the following way : Starting from $x$,
follow all \emph{unmatched} edges from $V1$ to the nodes in $V2$, and
from them all \emph{matched} edges back to $V1$ and so on.  For each
new node in $V1$ found (including $x$), increment in $1$ a variable
$k_1$, and for each new node in $V2$ do the same with $k_2$.  Both $k_1$
and $k_2$ start from zero.  Each new  node in $V2$ leads automatically
to a new node in $V1$, because the hypothesis implies that no exposed
node can be found.  Therefore when the BFS comes to an end (because all
$V2$ nodes have been visited once) we have that $k_1=k_2+1$ and we have
identified a subset of $k_1$ nodes of $V1$ adjacent to only $k_1 -1$
nodes of $V2$. $\hfill \diamond $ 
\\
Theorem~\ref{th:hask} means that, in order to recognize whether a given
graph is rigid, we have to count the number of independent bars it has,
or equivalently, be able to detect how many of them are dependent. Our
rigidity testing algorithm will be based upon \mbox{Theorem
\ref{th:equivalences}.D} , and consists in adding edges $e$ one at a
time to a set of independent edges $\hat E$, and testing whether this
enlarged $\hat E'$ is independent. If this is the case,  $e$  it is
definitively added to $\hat E$, otherwise $e$ is identified as a
dependent edge ($e$ is not independent of $\hat E$) and removed from the
graph.
\\
Adding a new edge $e$ to $\hat E$ produces the graph $G$, and its
associated bipartite graph $B(G)$. We know by
Theorem~\ref{th:equivalences}.D that $e$ is independent of $\hat E$ if
and only if a complete matching from $V1$ to $V2$ exists in $B(Gab)$
when any edge $ab$ in $\hat E' = e \cup \hat E$ is quadrupled.
Fortunately only the last edge $e$ needs be quadrupled, as the following
result due to Hendrickson~\cite{Hendrickson}  demonstrates.
\begin{lemma} 
\label{th:eonly}
Add a new edge $e$ to an independent edge set $\hat E$.  If a complete
matching in the sense of \mbox{theorem \ref{th:equivalences} } exists
when $e$ is quadrupled, then $\hat E'$ is independent.
\end{lemma}
{\it Proof:} assume $\hat E'$ is not independent. Then there
must exist some edge $e' \in \hat E'$ whose quadrupling causes
some subgraph $G'$ of $G$ to have ``too many edges''. This subgraph
\emph{must} include $e$ since $\hat E$ is an independent set. But this
subgraph would have the same number of edges if $e$ is quadrupled
instead of $e'$, therefore a complete matching would not be possible if
$e$ is quadrupled. \hfill $\diamond$
\\
Then in order to determine if a new edge $e$ is independent of a set
$\hat E$ we have to enlarge a bipartite matching to cover the 4 copies
of the new edge $e$ in $V1$. If this cannot be done, $e$ is redundant.
We have then an algorithm which enables us to identify redundant bars.
But we also need a means to identify rigid clusters in the system. The
advantage of the body-bar representation over the original formulation
in terms of joints is that each new rigid cluster which is identified
may be replaced by one node, therefore reducing the size of the system.
We will now see how these rigid clusters are detected. For this we need
the following results\cite{Hendrickson}.
\begin{generic} 
\label{th:only3}
If $3$ copies of a new edge $e$ are added to an independent set $\hat E$
generating a graph $G''$, then $B(G'')$ has a complete matching from
$V1$ to $V2$.
\end{generic}
{\bf Proof:} Assume there is no matching when $e$ is tripled. Then
there is some subgraph $\tilde G$ of $G''$ for which \mbox{$\tilde b > 3
\tilde n + 2 \tilde m$}. Remove the $3$ copies of $e$ and quadruple any
of the other edges. This graph has the same number of edges as $\tilde
G$ so it can also not be matched. But this is a contradiction if $\hat E$
is assumed to be independent. \hfill $\diamond$ 
\\
\begin{generic} 
\label{th:isostatic}
If the edge $e$ is dependent and $e$ is quadrupled, the failing
Hungarian tree spans a minimal subset of edges in $\hat E$ which form
an isostatic subgraph.
\end{generic}
{\bf Proof.}  Since we added 4 copies of the new edge, the number of
$v1$ nodes visited by the failing Hungarian tree is $b'+4$, where $b'$
of them belong to $E$.  The number of $v2$ nodes visited is by
construction of the bipartite graph \mbox{$3n'+2m'$.}  But in a previous
demonstration, we saw that when a BFS fails, it visits $k_1$ nodes in
$V1$ and $k_2$ nodes in $V2$ with \mbox{ $k_1=k_2+1$.}  Therefore, $b'+4
= 3n'+2m'+1$, or $b'=2m'+3n'-3$, which is the condition for an isostatic
subgraph. This subgraph is minimal, since removing any of its edges
would make the matching possible, by freeing one $v2$ node.  \hfill
$\diamond$ 
\\
The algorithm then proceeds as follows: Starting from a (possibly empty)
set $\hat E$ of independent edges, test new edges $e$ one by one by
adding them to $\hat E$ and trying to enlarge an already existing
matching of $B(G_e)$, which contains four copies of $e$ in $V1$.
\begin{itemize}
\item 
If all four copies can be matched, the new edge $e$ is independent.
Remove the three extra copies and add $e$ to the set of independent
edges $\hat E$. 
\item
If the matching fails for the fourth copy of $e$ (the first three can
always be matched), $e$ is dependent and the failing Hungarian
tree identifies a rigid subset of nodes. Remove all four copies.
\end{itemize}

\noindent
The rigid subgraph $G_a$  identified by a failed search may be
\emph{condensed} to a unique body $A$. This condensation step simply
amounts to a deletion of all bars and bodies visited during the failed
search, and replacing them by a unique body. Therefore the amount of
work involved in one condensation is equivalent to that needed for one
BFS.  All bars incident to $G_a$ are, after the condensation, incident
to body $A$. Condensation is only possible because we are able to
handle bodies, and is a key step in our algorithm.
\\
Bars whose both ends are found to be connected to the same rigid body
are not tested because they are obviously dependent.  This means that
the same subset of nodes is not condensed twice, so that each
condensation involves at least one node which has never been
condensed.  Therefore there will be at most $n$ condensations. The
condensation step requires just deleting all bars and bodies in the
failed Hungarian tree, therefore the total amount of work needed for
condensations is at most $O(n)$, since there are at most $O(n)$ bars in
$\hat E$, and no bar is condensed twice.
\\
A new bar leads either to a condensation or to a new element in $\hat
E$.  Both happen at most $O(n)$ times, so that at most $O(n)$ tests are
needed.  Since each test involves growing four Hungarian trees, each
one taking (at most) a time proportional to the number $O(n)$ of edges
in $\hat E$, the algorithm has a worst-case time complexity of
$O(n^2)$.
\\
This theoretical bound is the same as for Hendrickson's
algorithm~\cite{Hendrickson}. It is not difficult to understand why the
worst-case complexity of the body-bar algorithm cannot be better than
$O(n^2)$.  If no condensations ever occur, our algorithm is the same as
Hendrickson's. But this would only happen if there are no redundant
bars, since each redundant bar identifies a rigid subgraph and leads to
a condensation. We can see that while the joint-bar algorithm has its
worst-case performance in all cases in which there is long-range
rigidity~\cite{Note2}, the body-bar algorithm can only be pushed
towards $O(n^2)$ behavior in the improbable case of long-range rigidity
without redundancies (long-range isostatic rigidity). This situation is
not frequent in physical systems with disorder, where rigidity is
always redundant.  Therefore we expect the body-bar version here
introduced to perform much better than the original joint-bar version
on practical applications such as rigidity
percolation~\cite{JacobsThorpe,Letter},
glasses~\cite{Thorpe,Franzblau2} or granular materials~\cite{Guyon}.
\\
\section{ Implementation and performance of the algorithm.}
\label{sec:implementation}
The body-bar algorithm described in this work has been recently
applied~\cite{Letter} to study rigidity percolation on site
diluted-triangular lattices. In this work we will present the comparison
of performance between this algorithm and the original site-and-joint
version, for bond and site dilution on triangular lattices. Bonds or
sites are present on the lattice with probability $p$. Initially random
numbers are assigned to bonds in the following manner: In the bond
dilution case, each bond \texttt{ij} is assigned a random number
\texttt{brn(ij)}. For site dilution, sites \texttt{i} are given a random
number \texttt{srn(i)} and afterwards bonds are assigned
\hbox{\texttt{brn(ij) = max(srn(i),srn(j))}}. In both cases, bonds are
sorted in order of increasing \texttt{brn} and tested in that order. The
rest of the procedure is the same for bond or site dilution. This scheme
allows one to exactly detect the percolation point~\cite{Letter}, but is
not the only possible. For example, one could fix $p$ to a certain value
and, starting form an empty system, test all bonds for which
\texttt{brn} $< p$ in arbitrary order. The time-complexity of the
algorithm depends on the order in which bonds are tested, although its
results do not.
\\
The graph data structure is stored using based (pointer) variables
because in contrast to the regular lattice which originates it, the
multigraph has no regularity, and  its size changes during the
procedure, which makes static allocation of memory
impractical. A new bond is tested by adding four copies of it ($V1$
nodes) to the bipartite graph and attempting to match them to the
bodies ($3$ copies of each exist on the graph) or sites ($2$ copies of
each) in $V2$. If the four copies are matched, the new bond is marked
independent and its three additional copies are removed from the
graph.  If the fourth copy is not matched, the new bond is marked
redundant.  The set of bonds covered in the last search is in this case
a rigid subgraph. This rigid subgraph is \emph{minimal}, which means
that if any of its edges is removed then the matching would be
possible. This subgraph identifies then the subset of $E$ upon which
$e$ is dependent.  The concept of dependence can be recast to mean that
a self-stress is possible, therefore the failing BFS identifies the set
of edges of $G$ which would be stressed if the new edge $e$ is say, too
long or too short. This feature of the algorithm is very important in,
for example, rigidity percolation~\cite{Letter}, since it provides a
means to identify the stress-carrying part of a rigid cluster.
\\
Each time a rigid subgraph is identified a routine
\texttt{condensation} is called, which replaces all its elements
(enclosed bars and bodies) by a single body, putting three copies of it
in the graph. All external bars (bars incident to an enclosed body from
a non-enclosed body) are now incident to the new body. At a practical
level the only difference with Hendrickson's original algorithm is this
condensation step.  If the replacement of rigid objects by one node is
not done, one has the original bar-joint algorithm. One must in this
case~\cite{Hendrickson} mark all enclosed objects with the new rigid
label to avoid the need to test bonds whose both ends are connected to
the same rigid graph.
\\
As mentioned in Section~\ref{sec:laman}, multiple incidence joints must
be replaced by an auxiliary structure in order to have an equivalent
graph which is generic. In fact it is necessary to do this only in the
bond-diluted case, since for site dilution no multiple incidence points
are possible. For simplicity we always use auxiliary structures for
incidence points to bodies, although they are only necessary if the
number of incident bars is larger than two (Section~\ref{sec:laman}).
In this way the procedure is much simpler (otherwise the number of
incident bars should be checked after each condensation or addition of
a new edge) while the performance is not seriously affected.
\\ 
We saw already that rigid clusters will only be detected if a bond is
tested on them, that is, if a bond is found to be redundant. In other
words, the algorithm naturally identifies self-stressed (hyperstatic)
regions, while isostatic rigid clusters would go unnoticed if no bond
is tested on them. In the study of rigidity percolation~\cite{Letter},
one is interested in detecting the exact concentration $p_c$ of bonds
at which an isostatic rigid connection between two sides of the system
first appears. This is not automatically provided by the algorithm. A
possible way to do it would be the following: after testing each edge
$e$, test a \emph{fictitious} bond connecting opposite sides of the
sample. If the fictitious bond is independent, then these sides are not
rigidly connected. Otherwise the first time that the fictitious bond is
found dependent, a rigid connection is identified between the two sides
of the system. This method doubles the number of tests
(BFS's) needed.  We now describe a better option, which allows the
detection of the rigidity percolation point without extra
effort~\cite{Letter}.
\\
Two rigid bus-bars  are assumed to exist on the upper and lower edges
of the sample. These are represented as two  bodies $B_1$ and $B_2$,
and their corresponding three copies are set in $V_2$. Next a
fictitious bond $f$ connecting the two bus-bars is added to the graph.
A node $v_f \in V1$ represents this fictitious bond in the bipartite
graph $B(G)$, and is adjacent to three copies of $B_1$ and $B_2$. This
node $v_f$ (just one copy of it) is matched to one of these nodes
before starting to test any other edges. Next edges in the system are
tested in order of increasing random number \texttt{rn}. The first time
that an \emph{isostatic} rigid connection exists between the bus-bars,
because of the existence of this bond $f$ that already restricts one
relative degree of freedom between the busses, a dependent subgraph
will be found including the fictitious bond.  Thus the method to detect
isostatic percolation is simply checking, at each failed matching,
whether the fictitious bond $v_f$ has been visited during the last
search. If so, the last added edge $e$ is independent (the number
\texttt{rn} associated to this bond is $p_c$), and the subgraph visited
in the failed search is exactly the \emph{elastic
backbone}\footnote{The {\it spanning cluster} is the subset of edges
rigidly connected to both $B_1$ and $B_2$. The subset of it that would
be stressed if a pair of forces is applied between the busses is called
the {\it elastic backbone}.}.  All bars in this subgraph except $f$ are
\emph{cutting bonds}, that is, bonds whose individual removal would
produce the loss of rigidity (called \emph{red bonds} in scalar
percolation).  At the percolation point the fictitious bond is
removed.
\\

\leftline{\it Performance comparison.}
Both the body-bar and the bar-joint algorithms have a worst-case time
complexity that scales as $O(n^2)$.  But we argued that the body-bar
representation has a more convenient average-case behavior, since
searches are now done in a graph of much reduced number of elements whenever
there are rigidly connected subsets of sites. In
this section these differences are quantified. In order to do so we take
as a test case the randomly diluted triangular lattice. 
\begin{figure}[] \vbox{ 
\centerline{\psfig{figure=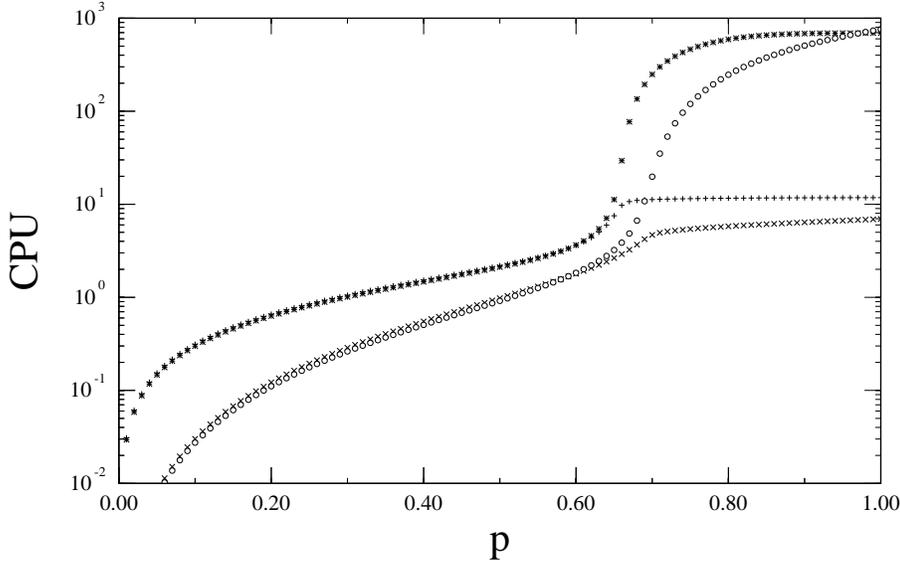,width=14cm} }
\caption{ Shown are total CPU times, in seconds, needed to test all
present bonds of a randomly diluted triangular lattice of size
$L=128$.  These measurements were done both on site- and bond-diluted
lattices, using the body-bar (($\times$): site dilution, ($+$): bond
dilution) and the joint-bar algorithms(($\circ$): site dilution, ($\ast$):
bond dilution) on a SPARC10 workstation. }
\label{fig:cmpCPU}
} \end{figure}
Total CPU times are measured as a function of $p$ for the bond- and
site-dilution cases.  Bonds are assigned a random number \texttt{rn} as
described in Section~\ref{sec:implementation}, and  tested sequentially
in order of increasing \texttt{rn}. In Figure~\ref{fig:cmpCPU} we see a
plot of CPU times required on typical lattice of linear size $L=128$
for both algorithms in the above mentioned cases. As expected, the
body-bar algorithm is more advantageous only in the rigid phase, since
the search for an exposed node takes place in a region which is
typically of the size of rigidly connected clusters. Each such region
is represented by one node in the body-bar case, and therefore searched
over in just one step. At larger scales, further collections of  bodies
are found to be rigidly  connected and therefore replaced by one body,
so that the size of the typical searches is kept almost $O(1)$.
\\
In Section~\ref{sec:intro} we mentioned that the time-complexity of
this procedure is of order $L^2$ (number of bonds tested) times the
size of the typical search for an augmenting path, which generally
scales as $L^\theta$, with $\theta$ a function of $p$. The overall CPU
time then scales as $L^\mu$ with $\mu=2+\theta(p)$. We estimate this
($p$-dependent) exponent by measuring CPU times for sizes $L=256, 128,
64$ and $32$. Averages were done over $10^2$ to $10^4$ samples.
Figure~\ref{fig:exponent} shows the value of the exponent $\mu$ for the
four cases under consideration, as a function of $p$. We see that the
body-bar algorithm has a time complexity that scales approximately as
$n^{1.12}$, while the bar-joint algorithm scales approximately as
$n^{1.9}$, which is not much better than the theoretical worst-case
limit $n^2$.
\begin{figure}[] \vbox{ 
\centerline{\psfig{figure=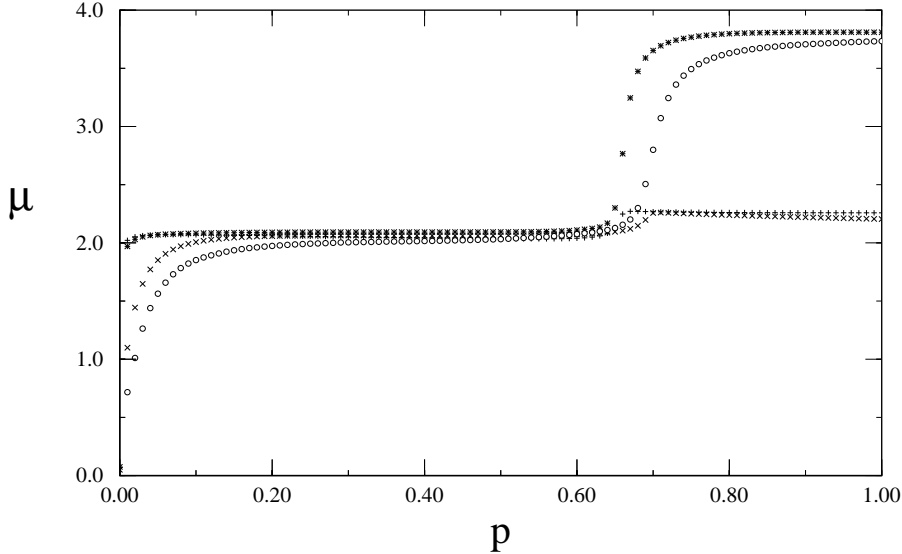,width=14cm} }
\caption{ CPU times as shown in Fig.~\ref{fig:cmpCPU} scale with system
size as $L^\mu$ with $\mu = 2 + \theta(p)$. The exponent $\theta(p)$
determines how the size of a typical search scales with size (see
text).  Symbols are the same as for Fig.~\ref{fig:cmpCPU}.
From the data in this plot we see that in the bar-joint algorithm, this
size grows almost as fast as $n$ in the rigid phase, while for the
body-bar algorithm described in this work this size remains almost
constant.  This improved behavior is due to ``condensation'' of rigidly
connected clusters in the body-bar algorithm.}
\label{fig:exponent}
} \end{figure}
%
\section{ Acknowledgments}
My interest in  the problem of rigidity was started by P.~M.~Duxbury,
and many of the ideas in this work were inspired by discussions with
him. An anonymous referee provided helpful advice concerning the
presentation of material in sections \ref{sec:intro} and
\ref{sec:rigmat}.  I thank the DOE under contract DE-FG02-90ER45418 and
the PRF for financial support. I also acknowledge support from the
Conselho Nacional de Desenvolvimento Cient\'\i fico e Tecnol\'ogico,
CNPq, Brazil.
\\
%
%
%

\end{document}